\documentclass[letterpaper, 10pt, conference]{ieeeconf}
\IEEEoverridecommandlockouts
\usepackage{amsmath,amssymb,amsfonts}
\usepackage{graphicx}
\usepackage{booktabs}
\usepackage{hyperref}
\usepackage{xcolor}
\usepackage{subcaption}
\usepackage{float}
\usepackage{cite}

\makeatletter
\renewcommand\@IEEEsectpunct{\textup{:}\ \,}
\makeatother

\title{\LARGE \bf Bilinear Input Modulation for Mamba: Koopman Bilinear Forms for Memory Retention and Multiplicative Computation}

\author{Hiroki Fujii and Masaki Yamakita
\thanks{\raggedright The authors are with Department of Systems and Control Engineering, Institute of Science Tokyo, Tokyo, Japan. E-mail: \texttt{fujii.h.8f58@m.isct.ac.jp}, \texttt{yamakita@ac.ctrl.titech.ac.jp}.}
}

\begin{document}
\maketitle

\begin{abstract}
Selective State Space Models (SSMs), notably Mamba, employ diagonal state transitions that limit both memory retention and bilinear computational capacity.
We propose a factorized bilinear input modulation that augments the SSM with a state-input product, interpretable as a finite-dimensional Koopman bilinear form.
After introducing a shared state across channels (Coupled SSM), the modulation admits three implementations. Coupled Bilinear Input Modulation (seq-BIM) retains the full bilinear product on the input side at the cost of sequential computation, Coupled Gated Modulation (GM) linearizes it into a gate modulation that is compatible with the parallel scan, and Parallel Bilinear Input Modulation (p-BIM) places the same bilinear product on the state transition while remaining parallel-scannable.
Experiments on a multiple input-delay pendulum (memory retention) and NARMA-10 (bilinear computation) reveal a clear dissociation. GM substantially improves memory retention but not bilinear computation, while both seq-BIM and p-BIM improve both.
A pathway ablation confirms that the two downstream routes of the bilinear signal serve complementary roles.
The improvement is statistically robust, with the bilinear variants consistently outperforming the other variants on bilinear computation.
Furthermore, only the bilinear variants benefit from increasing the SSM state dimension, while coupling or gate modulation alone show no improvement, establishing the bilinear mechanism as uniquely capable of exploiting larger state spaces.
\end{abstract}


\section{Introduction}

While classical recurrent neural networks such as Echo State Networks~\cite{jaeger2001echo} use nonlinear activations to implicitly mix input history, Selective State Space Models (SSMs), building on the HiPPO framework~\cite{gu2020hippo} and structured state spaces (S4)~\cite{gu2022efficiently}, have evolved into efficient alternatives to Transformers for sequence modeling by replacing nonlinear recurrences with structured linear ones.
Mamba~\cite{gu2023mamba} introduced input-dependent selectivity, and Mamba-2~\cite{dao2024transformers} unified the SSM perspective with linear attention via the state space duality (SSD) framework.
Recently, Wang et al.~\cite{wang2025ispm} demonstrated that Mamba-based architectures can achieve accurate prediction of chaotic dynamical systems under teacher forcing, establishing the Integrated State Prediction Model (ISPM) framework that our work builds upon.

At their core, these SSMs maintain a hidden state $h_t$ that evolves via a \emph{linear} recurrence $h_t = A_t h_{t-1} + B_t x_t$, where the diagonal transition matrix $A_t$ and input matrix $B_t$ are input-dependent.
This linearity enables parallel training via associative scan, and diagonal state matrices have been shown to be as effective as structured ones~\cite{gupta2022diagonal,gu2022parameterization}. However, the diagonal linear recurrence fundamentally limits representational power, as no multiplicative interaction between the hidden state and the input can occur within the recurrence itself.
Recent work has improved expressiveness while preserving linearity. Mamba-3~\cite{lahoti2026mamba3} introduces a higher-order discretization and multi-input aggregation into a shared state, achieving greater state efficiency within the linear recurrence.
However, these advances do not address the absence of state-input products, which are essential for representing bilinear dynamics.

Koopman operator theory~\cite{koopman1931hamiltonian,brunton2022modern} provides a principled framework for embedding nonlinear dynamical systems into linear evolution equations.
For an autonomous system $z_{t+1} = F(z_t)$, the Koopman operator $\mathcal{K}$ acts on \emph{observables} $g(z)$ such that $g(z_{t+1}) = \mathcal{K} g(z_t)$, lifting the nonlinear dynamics into a linear space.
When control inputs $u_t$ are present, the extended Koopman framework yields a \emph{bilinear} form~\cite{proctor2018generalizing,surana2016koopman}:
\begin{equation}
  g(z_{t+1}) = A\, g(z_t) + B\, u_t + \sum_j N_j\, g(z_t)\, u_t^{(j)}
  \label{eq:koopman_bilinear}
\end{equation}
where the bilinear terms $N_j g(z_t) u_t^{(j)}$ capture the state-input interaction that a purely linear model cannot represent.
Bruder et al.~\cite{bruder2021advantages} demonstrate that bilinear Koopman realizations consistently improve with additional basis functions, and this framework has been applied to model predictive control~\cite{kanai2023lifted} and power grid prediction~\cite{jiang2024modularized}.

We apply this bilinear perspective to the Mamba architecture. After introducing a shared state across channels (Coupled SSM), we derive three modulation variants. Coupled Bilinear Input Modulation (seq-BIM) retains the full state-input product on the input side but requires sequential computation, and Coupled Gated Modulation (GM) freezes $h$ at a learned prior to recover parallelizability. The parallelizability-accuracy trade-off motivates an alternative placement. Parallel Bilinear Input Modulation (p-BIM) puts the same bilinear product on the state transition, keeping the full matrix-valued product while remaining parallel-scannable. Because bilinear modulation introduces a cross-term between the input $x$ and the state $h$, we refer to the resulting architecture as Mamba-X (Mamba-Cross).
Experiments on a multiple input-delay pendulum and NARMA-10 reveal a clear dissociation. GM improves memory retention, while both seq-BIM and p-BIM additionally enable bilinear computation.

\section{Architecture: From Diagonal SSM to Bilinear Input Modulation}

We describe the progression from Standard Mamba through three successive modifications, each adding a specific structural capability.

\subsection{Step 0: Standard Mamba (Diagonal SSM)}

In Standard Mamba, the SSM branch processes a $d_{\mathrm{inner}}$ (hereafter $d_i$)-dimensional input $x_t$ through $d_i$ independent channels, each maintaining its own $d_{\mathrm{state}}$ (hereafter $d_s$)-dimensional state.
The state update for channel~$d \in \{1,\ldots,d_i\}$ and state dimension~$n \in \{1,\ldots,d_s\}$ is:
\begin{equation}
  h_t^{(d,n)} = \mathrm{dA}_t^{(d,n)} \cdot h_{t-1}^{(d,n)} + \Delta t_t^{(d)} \cdot B_t^{(n)} \cdot x_t^{(d)}
  \label{eq:standard}
\end{equation}
where $\mathrm{dA}_t^{(d,n)} = e^{A^{(d,n)} \cdot \Delta t_t^{(d)}}$ is the discretized decay, $A_{\log} \in \mathbb{R}^{d_i \times d_s}$ parameterizes the decay rates ($A^{(d,n)} = -\exp(A_{\log}^{(d,n)}) < 0$), and $\Delta t_t$, $B_t$, $C_t$ are input-dependent selectivity parameters produced by a learned linear projection $x_{\mathrm{proj}}$ applied to $x_t$. The projection output is split into the discretization step $\Delta t_t$ (via softplus), the input mixing $B_t$, and the readout $C_t$.
The output is:
\begin{equation}
  y_t^{(d)} = \sum_{n} C_t^{(n)} \cdot h_t^{(d,n)} + D^{(d)} \cdot x_t^{(d)}
\end{equation}
where $D \in \mathbb{R}^{d_i}$ is a learnable skip-connection weight.
The total state comprises $(d_i \times d_s)$ scalars, all evolving independently.

\subsection{Step 1: Coupled SSM -- Shared State with Input/Output Coupling}
\label{sec:coupled}

The first modification collapses the $d_i$ independent channels into a single shared state vector $h_t \in \mathbb{R}^{d_s}$, and introduces dense coupling matrices to mix information between the input space and the shared state:
\begin{align}
  x_{\mathrm{state},t} &= B_{\mathrm{coup}}\, x_t
  \label{eq:coupled_in} \\
  h_t &= \mathrm{dA}_t \odot h_{t-1} + \Delta t_t \odot B_t \odot x_{\mathrm{state},t}
  \label{eq:coupled_state} \\
  y_t &= C_{\mathrm{coup}}\,(C_t \odot h_t) + D \odot x_t
  \label{eq:coupled_out}
\end{align}
where $\mathrm{dA}_t = e^{A \cdot \Delta t_t}$ with shared diagonal $A \!\in\! \mathbb{R}^{d_s}$, $B_{\mathrm{coup}} \!\in\! \mathbb{R}^{d_s \times d_i}$, and $C_{\mathrm{coup}} \!\in\! \mathbb{R}^{d_i \times d_s}$.

$B_{\mathrm{coup}}$ projects the $d_i$-dimensional input into the $d_s$-dimensional state space, and $C_{\mathrm{coup}}$ projects back.
The diagonal $A$ is now shared across all channels.
This creates inter-channel communication through the shared state (channel $d$ can influence channel $d'$ via $B_{\mathrm{coup}}$ and $C_{\mathrm{coup}}$), but the recurrence in Eq.~\eqref{eq:coupled_state} remains linear in $h$, preserving compatibility with the parallel scan algorithm.

\subsection{Steps 2A \& 3: Bilinear Input Modulation on the Input Side}
\label{sec:coupled_p1}

The Coupled SSM of Section~\ref{sec:coupled} corresponds to a Koopman model with only the linear terms $A g(z)$ and $B u$ in Eq.~\eqref{eq:koopman_bilinear}.
To capture the bilinear interaction $N_j g(z) u^{(j)}$, we add a factorized bilinear input modulation that modulates the input using the hidden state before the SSM update.

\subsubsection{Step 2A: seq-BIM -- Full bilinear modulation on the input side}
\label{sec:seq_bim}

At each timestep, the input is modified as:
\begin{align}
  h_{\mathrm{proj}} &= \tanh\!\big(W_h\, h_{t-1} \cdot s\big) \label{eq:p1_hproj} \\
  x_{\mathrm{mod},t} &= x_t + W_{\mathrm{out}}\!\big((W_x\, x_t) \odot h_{\mathrm{proj}}\big)
  \label{eq:p1_xmod}
\end{align}
where $W_h \in \mathbb{R}^{d_i \times d_s}$, $W_x, W_{\mathrm{out}} \in \mathbb{R}^{d_i \times d_i}$, and $s = 1/\sqrt{d_i}$.
The $\tanh$ bounds $h_{\mathrm{proj}}$ to $[-1,1]$, preventing the bilinear product from amplifying large hidden states and stabilizing training.

The modulated input $x_{\mathrm{mod},t}$ then replaces $x_t$ in the entire Coupled SSM pipeline. It is fed through the selectivity projection $x_{\mathrm{proj}}$ to produce $\Delta t_t$, $B_t$, $C_t$, and through $B_{\mathrm{coup}}$ to produce $x_{\mathrm{state},t}$.
The state update and output follow Eqs.~\eqref{eq:coupled_in}--\eqref{eq:coupled_out} with $x_t$ replaced by $x_{\mathrm{mod},t}$.

The critical term is the element-wise product $(W_x x_t) \odot h_{\mathrm{proj}}$ in Eq.~\eqref{eq:p1_xmod}. This is a factorized approximation of the Koopman bilinear term $N_j g(z) u^{(j)}$, where the hidden state $h$ plays the role of the Koopman observable $g(z)$ and $x_t$ plays the role of the control input~$u$.
The factorization through $W_x$, $W_h$, and $W_{\mathrm{out}}$ decomposes the full bilinear tensor into a diagonal Hadamard product followed by a linear projection, keeping the parameter cost at $O(d_i^2)$ rather than the $O(d_i^2 \cdot d_s)$ of an explicit bilinear tensor.
Because $x_{\mathrm{mod}}$ flows through $x_{\mathrm{proj}}$, this bilinear term affects all three selectivity parameters ($\Delta t$, $B$, $C$) simultaneously, enabling the model to perform multiplicative computation over its memory contents.

However, since $x_{\mathrm{mod},t}$ depends on $h_{t-1}$ through Eq.~\eqref{eq:p1_hproj}, the transition is no longer linear in~$h$.
This breaks the associative property required for parallel scan, forcing sequential computation with $O(L)$ depth.

\subsubsection{Step 3: GM -- Linearising the $h$-dependency for parallel scan}
\label{sec:linear_p1}

To recover parallelizability, we derive a variant of seq-BIM that is linear in $h$.
Dropping the $\tanh$ in Eq.~\eqref{eq:p1_hproj} and absorbing the scaling $s$ into the linear map, the modulation in Eq.~\eqref{eq:p1_xmod} becomes linear in $h$:
\begin{align}
  x_{\mathrm{mod},t} &= x_t + W_{\mathrm{out}}\!\big((W_x\, x_t) \odot (s \cdot W_h\, h_{t-1})\big) \nonumber \\
                      &= x_t + M(x_t)\, h_{t-1}
  \label{eq:lp1_linearize}
\end{align}
where $M(x_t) = s \cdot W_{\mathrm{out}} \cdot \mathrm{diag}(W_x\, x_t) \cdot W_h \in \mathbb{R}^{d_i \times d_s}$ is a matrix that depends on $x_t$ but not on $h$.

\paragraph{Why the modulation appears in the gate}
In seq-BIM, $x_{\mathrm{mod}}$ flows into two downstream pathways: (1)~$B_{\mathrm{coup}}$ for state input, and (2)~$x_{\mathrm{proj}}$ for selectivity parameters.
We first trace pathway~(1).
Substituting Eq.~\eqref{eq:lp1_linearize} into the state input $x_{\mathrm{state}} = B_{\mathrm{coup}}\, x_{\mathrm{mod}}$ and defining $G(x_t) = B_{\mathrm{coup}}\, M(x_t) \in \mathbb{R}^{d_s \times d_s}$:
\begin{equation}
  x_{\mathrm{state},t} = B_{\mathrm{coup}}\, x_t + G(x_t)\, h_{t-1}
\end{equation}
Inserting into the state update (Eq.~\eqref{eq:coupled_state}):
\begin{align}
  h_t &= \mathrm{dA}_t \odot h_{t-1} + \Delta t_t \odot B_t \odot \big(B_{\mathrm{coup}}\, x_t + G(x_t)\, h_{t-1}\big) \nonumber \\
      &= \big(\mathrm{dA}_t + \Delta t_t \odot B_t \odot G(x_t)\big) \odot h_{t-1} \nonumber \\
      &\quad + \Delta t_t \odot B_t \odot B_{\mathrm{coup}}\, x_t
  \label{eq:lp1_derivation}
\end{align}
The $B_{\mathrm{coup}} M(x_t) h_{t-1}$ term contains $h_{t-1}$, so it algebraically migrates to the gate (the coefficient of $h_{t-1}$), while the input term $\Delta t_t \odot B_t \odot B_{\mathrm{coup}} x_t$ remains unmodified.
This is not a design choice but a mathematical consequence of the linearization.

\paragraph{The selectivity pathway is lost}
Pathway~(2), $x_{\mathrm{mod}} \to x_{\mathrm{proj}} \to (\Delta t, B_t, C_t)$, would modify $\Delta t_t$ and $B_t$ themselves.
However, $\Delta t = \mathrm{softplus}(W_{\Delta}\, x_{\mathrm{mod}})$ is nonlinear in $x_{\mathrm{mod}}$, and linearizing this pathway with respect to $h$ is intractable in closed form.
GM therefore drops the selectivity pathway entirely, retaining only the gate modulation from the $B_{\mathrm{coup}}$ pathway.
As we show in Section~\ref{sec:ablation}, this omission explains why GM underperforms on NARMA-10. The selectivity pathway is precisely what enables bilinear computation over memory.

\paragraph{Stabilization via sigmoid}
The derived gate $\mathrm{dA}_t + \Delta t_t \odot B_t \odot G(x_t)$ in Eq.~\eqref{eq:lp1_derivation} is not guaranteed to remain in $(0,1)$. The modulation can push it above 1 (causing divergence) or below 0.
We stabilize by replacing the additive gate with a sigmoid, where $n \in \{1, \ldots, d_s\}$ indexes the state dimension:
\begin{align}
  g_t^{(n)} &= \sum_d B_{\mathrm{coup}}^{(n,d)} \cdot \big[W_{\mathrm{out}}\!\big((W_x\, x_t) \odot W_h^{(:,n)}\big)\big]_d
  \label{eq:lp1_gatemod} \\
  \mathrm{gate}_t^{(n)} &= \sigma\!\Big(\underbrace{A^{(n)} \cdot \Delta t_t^{(n)}}_{\log \mathrm{dA}_t^{(n)}} + \underbrace{\Delta t_t^{(n)} \cdot B_t^{(n)} \cdot g_t^{(n)} \cdot s}_{\text{GM modulation}}\Big)
  \label{eq:lp1_gate}
\end{align}
where $W_h^{(:,n)}$ denotes the $n$-th column of $W_h$, a learned constant vector that replaces the dynamic $h_{t-1}$ of seq-BIM.
This can be interpreted as freezing $h$ at a learned prior for each state dimension, rather than at a dynamic runtime value.
When $g_t^{(n)} > 0$, the gate shifts toward~1 (slower decay, improved retention), and when $g_t^{(n)} < 0$, toward~0 (faster forgetting).

\paragraph{Parallel scan compatibility}
Because $g_t^{(n)}$ depends only on $x_t$ and not on $h_{t-1}$, the modified recurrence
\begin{equation}
  h_t^{(n)} = \mathrm{gate}_t^{(n)} \cdot h_{t-1}^{(n)} + \Delta t_t^{(n)} \cdot B_t^{(n)} \cdot x_{\mathrm{state},t}^{(n)}
\end{equation}
is linear in $h$ and fully compatible with the parallel scan algorithm ($O(\log L)$ depth).

\subsection{Step 2B: Bilinear Input Modulation on the $A$-Side (p-BIM)}
\label{subsec:pbim}

An alternative bilinear placement branches directly off of Step 1
(Coupled-SSM): rather than routing the bilinear product through the
input $x_t$ as seq-BIM does, place it directly on the \emph{state
transition} (the $A$ side). The resulting variant keeps the full
bilinear product of seq-BIM while admitting a single associative
scan over the sequence.
Using the same factorised $(W_h, W_x, W_{\mathrm{out}})$ parameterisation
as seq-BIM, we build a full $(d_s, d_s)$ matrix
\begin{align}
  \label{eq:pbim_M}
  M(x_t) &= \tfrac{1}{\sqrt{d_i}}\, W_{\mathrm{out}}
            \operatorname{diag}(W_x x_t)\, W_h, \\
  \label{eq:pbim_N}
  N(x_t) &= (\Delta t_t \odot B_t) \odot_{\mathrm{row}}
            \big[ B_{\mathrm{coup}}\, M(x_t)\big],
\end{align}
where $u \odot_{\mathrm{row}} M$ denotes the row-wise scaling of matrix
$M \in \mathbb{R}^{d_s \times d_s}$ by vector $u \in \mathbb{R}^{d_s}$, so that
$(u \odot_{\mathrm{row}} M)_{n,m} = u_n M_{n,m}$.
We use $N(x_t)$ as an additive correction to the diagonal Mamba gate, defining the matrix-valued gate $G(x_t) = \operatorname{diag}(e^{A \Delta t_t}) + N(x_t)$:
\begin{multline}
  \label{eq:pbim_gate}
  h_{t+1} = G(x_t)\, h_t + \Delta t_t \odot B_t \odot (B_{\mathrm{coup}} x_t).
\end{multline}
Because $G(x_t)$ depends only on $x_t$ and not on $h$, the associative scan combiner $(G_2, b_2)\oplus(G_1, b_1) = (G_2 G_1,\, G_2 b_1 + b_2)$ is well defined and p-BIM is parallel-scan compatible.

\begin{table}[t]
\centering
\scriptsize
\setlength{\tabcolsep}{1.5pt}
\begin{tabular}{@{}l@{\hspace{1pt}}ccccc@{}}
\toprule
\textbf{Property} & \textbf{Std} & \textbf{Coup} & \textbf{seq-BIM} & \textbf{GM} & \textbf{p-BIM} \\
\midrule
State shape       & $(d_i, d_s)$ & $(d_s,)$   & $(d_s,)$ & $(d_s,)$ & $(d_s,)$ \\
$B/C_{\mathrm{coup}}$ & ---       & \checkmark & \checkmark & \checkmark & \checkmark \\
Modulation        & ---          & ---        & $W_h, W_x, W_{\mathrm{out}}$ & $W_h, W_x, W_{\mathrm{out}}$ & $W_h, W_x, W_{\mathrm{out}}$ \\
Target            & ---          & ---        & Input $x_t$ & Gate $\mathrm{dA}_t$ & $N(x_t)$ \\
Parallelizable    & \checkmark   & \checkmark & \texttimes & \checkmark & \checkmark \\
\bottomrule
\end{tabular}
\caption{Architecture progression. Standard Mamba, Coupled SSM (adds $B_{\mathrm{coup}}/C_{\mathrm{coup}}$), and the three modulated variants (seq-BIM, GM, p-BIM).}
\label{tab:arch}
\end{table}

\section{Experiments}

\subsection{Multiple Input-Delay Pendulum: Isolating Memory Retention}

The input-delay task simulates a damped pendulum driven by a finite impulse response (FIR) buffer of past inputs:
\begin{align}
  \omega_{t+1} &= \omega_t + \Big[-\tfrac{g}{\ell}\sin\theta_t - b\,\omega_t + \sum_{k=0}^{K-1} w_k\, u_{t-k}\Big]\,\Delta t \\
  \theta_{t+1} &= \theta_t + \omega_{t+1}\,\Delta t
\end{align}
with normalized weights $w_k = e^{-\gamma k} / \sum_{j=0}^{K-1} e^{-\gamma j}$ (decay factor $\gamma{=}0.15$, $K{=}24$, $b{=}0.05$, $\Delta t{=}0.01$).
The model must retain a sliding window of $K$ past input values.
This is a \emph{linear memory} task, where past inputs are combined linearly through the FIR convolution, and no products between stored values are required.

\subsection{NARMA-10: Requiring Nonlinear Computation over Memory}

The NARMA-10 (Nonlinear AutoRegressive Moving Average) benchmark, introduced by Atiya and Parlos~\cite{atiya2000new}:
\begin{equation}
  y_{t+1} = 0.3\,y_t + 0.05\,y_t\!\sum_{i=0}^{9} y_{t-i} + 1.5\,u_{t-9}\,u_t + 0.1
\end{equation}
contains two bilinear terms:
(1)~$y_t \cdot \sum y_{t-i}$, a product of current state with a memory sum, and
(2)~$u_{t-9} \cdot u_t$, a product of current input with a delayed input.
In addition to the memory retention required by the input-delay task, the main challenge lies in retrieving $u_{t-9}$ specifically from past states.

\subsection{Setup}
\label{sec:setup}

Our implementation builds on the Integrated State Prediction Model (ISPM)~\cite{wang2025ispm}, which uses Mamba for prediction of dynamical systems.
All proposed modifications (coupling matrices, seq-BIM, p-BIM, GM) are applied exclusively to the Mamba block within the ISPM framework. The surrounding pipeline (input projection, convolution, gating, output projection) remains unchanged.

Unless explicitly specified, all models use $d_s{=}8$ and $d_i{=}d_{\mathrm{model}} {\times} 4$ (where $d_{\mathrm{model}}$ is the input dimension, 3 for the input-delay task and 2 for NARMA-10).
Training data consists of 66,000 trajectories, each with an input context length of $L{=}50$ steps, and test data consists of 5,000 trajectories of the same length.
All models are trained for 200K iterations with Adam (batch size 100, learning rate $10^{-3} \to 10^{-5}$ via cosine annealing).
Both benchmarks use 11 seeds for the main results.
Section~\ref{app:scaling} explores the effect of varying $L$ and $d_s$ beyond these defaults.
We compare five architectures described in Section~II: Standard Mamba, Coupled, GM, seq-BIM, and p-BIM (Table~\ref{tab:arch}).

Evaluation is based on autoregressive (AR) rollout MSE on 100 held-out trajectories of 250 steps each. The first $L{-}1$ steps use teacher forcing (TF) as warmup, and the remaining steps are predicted autoregressively. AR MSE is computed over the autoregressive portion only.
On the input-delay task, the bilinear weights $W_h, W_x, W_{\mathrm{out}}$ are initialized with a Gaussian of std $0.5$, which empirically outperforms smaller values on this benchmark. A principled analysis is left to future work.

\section{Results}

\subsection{Input-Delay Task: Memory Retention}

\begin{table}[t]
\centering
\footnotesize
\begin{tabular}{lccccc}
\toprule
\textbf{Model} & \textbf{Mean} & \textbf{Med.} & \textbf{Worst} & \textbf{SD} & \textbf{Impr.} \\
\midrule
Standard          & 16.99 & 2.420 & 106.6 & 31.16 & $1.0\times$ \\
Coupled           & 27.05 & 5.928 & 171.1 & 51.97 & $0.6\times$ \\
GM        & 0.692 & 0.760 & \textbf{1.28} & \textbf{0.44} & $24.5\times$ \\
seq-BIM   & \textbf{0.317} & \textbf{0.160} & 1.78 & 0.50 & $\mathbf{53.6\times}$ \\
p-BIM$^{\dagger}$     & 0.662 & 0.463 & 2.21 & 0.64 & $25.7\times$ \\
\bottomrule
\end{tabular}
\caption{Input-delay results ($K{=}24$, $n_{\mathrm{steps}}{=}50$, 11 seeds). \textit{SD} stands for standard deviation. \textit{Impr.}\ is the mean improvement over Standard. $\dagger$ p-BIM training diverges on 1 of the 11 seeds, and the reported statistics are aggregated over the remaining 10 seeds.}
\label{tab:ringbuffer}
\end{table}

\begin{figure}[t]
\centering
\includegraphics[width=\columnwidth]{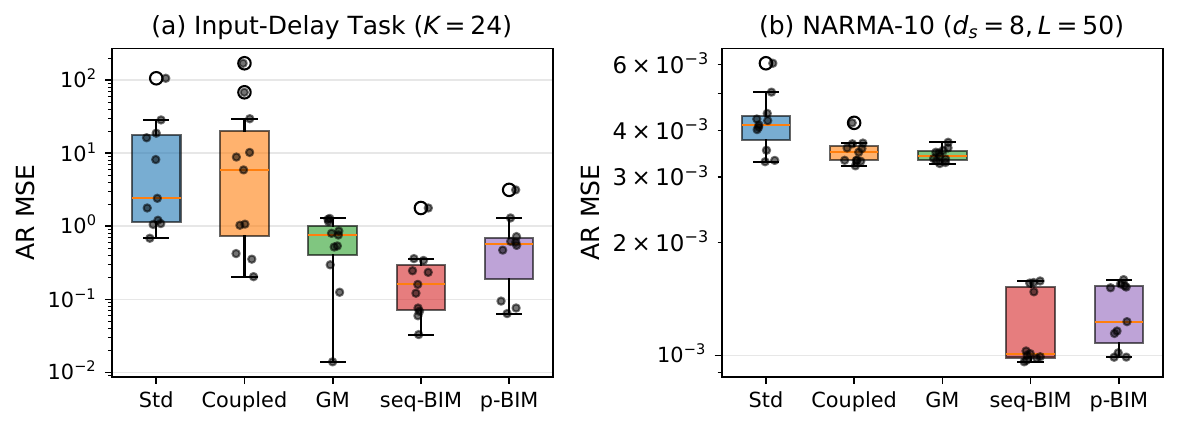}
\caption{AR MSE distribution over 11 seeds. (a) Input-delay task. All three modulated variants (GM, seq-BIM, p-BIM) achieve lower MSE, with seq-BIM the lowest, while Standard and Coupled have high variance. (b) NARMA-10. Both bilinear variants (seq-BIM, p-BIM) separate from Standard, Coupled, and GM by a clear gap, and they differ from each other by less than the seed variance.}
\label{fig:box_combined}
\end{figure}

\begin{figure}[t]
\centering
\includegraphics[width=\columnwidth]{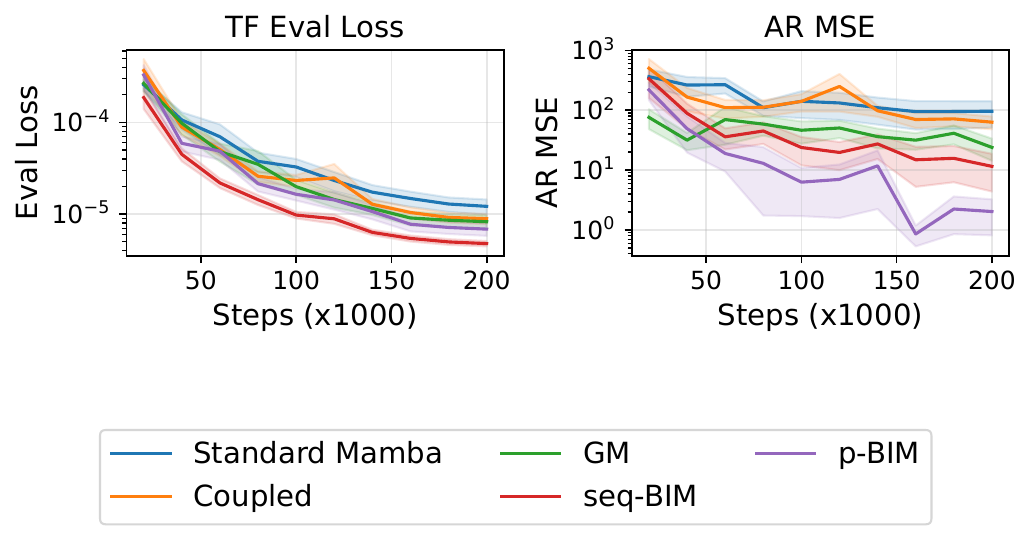}
\caption{Input-delay task: Training loss curves (11 seeds, mean $\pm$ SEM). seq-BIM achieves the lowest AR MSE.}
\label{fig:rb_loss}
\end{figure}

All three modulated variants achieve more than $20\times$ mean improvement over Standard with 11 seeds (Table~\ref{tab:ringbuffer}, Fig.~\ref{fig:box_combined}a, Fig.~\ref{fig:rb_loss}).
seq-BIM achieves the lowest median AR MSE (0.160), followed by p-BIM (0.463) and GM (0.760). Both GM and p-BIM are fully parallelizable, with p-BIM additionally retaining the full matrix-valued bilinear product.
Coupled provides no improvement, confirming that dense coupling alone is insufficient.
While p-BIM reaches the seq-BIM accuracy regime on the majority of seeds, the across-seed variance raises its mean above the seq-BIM baseline. The worst convergent seed attains AR MSE $2.21$, and training diverges on one of the eleven seeds.
Unlike GM, where the sigmoid keeps the diagonal gate in $(0,1)$ by construction, the matrix-valued gate $G(x_t)$ of p-BIM (\S\ref{subsec:pbim}) carries no spectral bound on $N(x_t)$, allowing the gate to grow during training.
Stabilising p-BIM across all seeds for the input-delay task should be addressed in the future work.

\subsection{NARMA-10: Nonlinear Computation}

\begin{table}[t]
\centering
\scriptsize
\setlength{\tabcolsep}{3pt}
\begin{tabular}{@{}lccccc@{}}
\toprule
\textbf{Model} & \textbf{Mean} & \textbf{Med.} & \textbf{Worst} & \textbf{SD} & \textbf{Impr.} \\
\midrule
Standard         & $4.22\mathrm{e}{-3}$ & $4.13\mathrm{e}{-3}$ & $6.04\mathrm{e}{-3}$ & $7.5\mathrm{e}{-4}$ & $1.0\times$ \\
Coupled          & $3.52\mathrm{e}{-3}$ & $3.49\mathrm{e}{-3}$ & $4.18\mathrm{e}{-3}$ & $2.6\mathrm{e}{-4}$ & $1.2\times$ \\
GM       & $3.43\mathrm{e}{-3}$ & $3.41\mathrm{e}{-3}$ & $3.71\mathrm{e}{-3}$ & $1.4\mathrm{e}{-4}$ & $1.2\times$ \\
seq-BIM  & $\mathbf{1.19\mathrm{e}{-3}}$ & $\mathbf{1.01\mathrm{e}{-3}}$ & $1.58\mathrm{e}{-3}$ & $2.7\mathrm{e}{-4}$ & $\mathbf{3.5\times}$ \\
p-BIM    & $1.30\mathrm{e}{-3}$ & $1.23\mathrm{e}{-3}$ & $1.59\mathrm{e}{-3}$ & $2.5\mathrm{e}{-4}$ & $3.3\times$ \\
\bottomrule
\end{tabular}
\caption{NARMA-10 results (11 seeds, 200K iterations). seq-BIM achieves $3.5\times$ improvement over Standard, while the parallel-scannable p-BIM matches it within noise at $3.3\times$.}
\label{tab:narma}
\end{table}

\begin{figure}[t]
\centering
\includegraphics[width=\columnwidth]{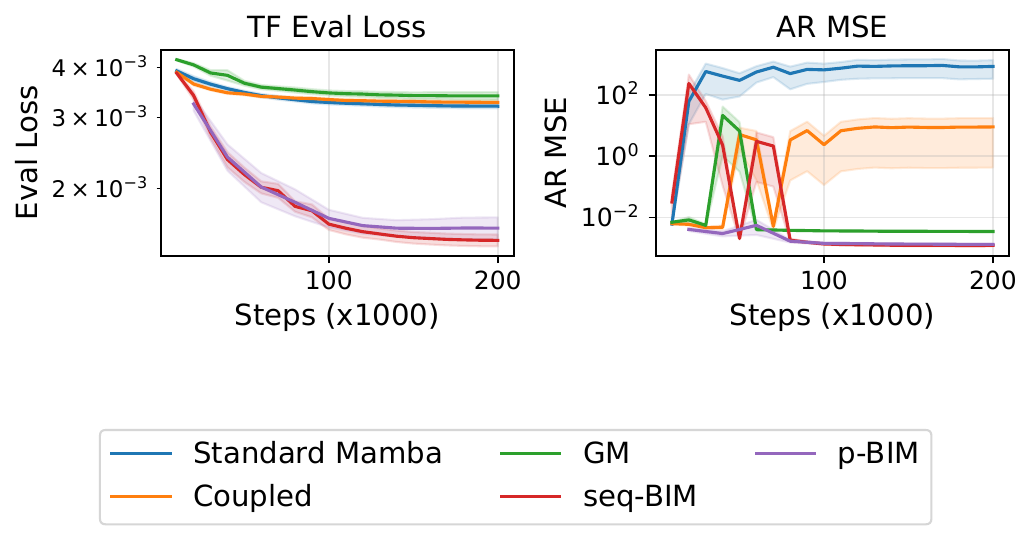}
\caption{NARMA-10: Training loss curves (11 seeds, mean $\pm$ SEM). Both bilinear variants (seq-BIM and p-BIM) separate clearly in TF Eval loss from the non-bilinear baselines, and the two bilinear curves nearly coincide. Standard and Coupled diverge in AR MSE.}
\label{fig:narma_loss}
\end{figure}

Both seq-BIM and p-BIM separate from the non-bilinear
baselines on NARMA-10 (Table~\ref{tab:narma}, Fig.~\ref{fig:box_combined}b),
achieving $3.5\times$ and $3.3\times$ improvement respectively, while
GM and Coupled gain only $1.2\times$ each. With 11 seeds, the
separation is statistically robust; the two bilinear variants differ by
less than 10\% in mean, within seed variance.
The training loss curves (Fig.~\ref{fig:narma_loss}) show that both
bilinear variants separate from the non-bilinear models at
$\sim$50K steps.

The advantage of both bilinear variants grows further when the shared state dimension is increased. At $d_s{=}16$, seq-BIM achieves over $70\times$ improvement on NARMA-10 over Standard and p-BIM reaches over $90\times$, while Standard, Coupled, and GM gain little from increasing $d_s$ (Section~\ref{app:scaling}, Table~\ref{tab:arch_sep}, Fig.~\ref{fig:l_sweep}).
Both follow the same $d_s$-scaling regardless of placement, so the gain comes from the bilinear mechanism rather than its position (Fig.~\ref{fig:l_sweep}).

\subsection{Pathway Ablation: Where Does seq-BIM Act?}
\label{sec:ablation}

In seq-BIM, the modulated input $x_{\mathrm{mod}}$ is used in two downstream pathways:
\begin{enumerate}
  \item \textbf{Selectivity pathway} ($x_{\mathrm{mod}} \to x_{\mathrm{proj}} \to (\Delta t, B_t, C_t)$)
  \item \textbf{State input pathway} ($x_{\mathrm{mod}} \to B_{\mathrm{coup}} \to x_{\mathrm{state}}$)
\end{enumerate}
To isolate the contribution of each pathway, we construct two ablation variants that route $x_{\mathrm{mod}}$ to only one pathway while using the original $x_t$ for the other.

\begin{table}[t]
\centering
\footnotesize
\begin{tabular}{lcccc}
\toprule
\textbf{Variant} & \textbf{$x_{\mathrm{proj}}$} & \textbf{$B_{\mathrm{coup}}$} & \textbf{In-Delay} & \textbf{NARMA} \\
\midrule
seq-BIM (full) & $x_{\mathrm{mod}}$ & $x_{\mathrm{mod}}$ & 0.667 & $1.54\mathrm{e}{-3}$ \\
xproj-only   & $x_{\mathrm{mod}}$ & $x_t$              & 0.259          & $1.54\mathrm{e}{-3}$ \\
bcoup-only   & $x_t$              & $x_{\mathrm{mod}}$ & 26.47          & $1.96\mathrm{e}{-3}$ \\
\midrule
Coupled   & $x_t$ & $x_t$ & 60.12 & $3.70\mathrm{e}{-3}$ \\
\bottomrule
\end{tabular}
\caption{Pathway ablation (3 seeds). On the input-delay task, the selectivity pathway (xproj-only) has a greater influence. On NARMA-10, xproj-only matches the full model.}
\label{tab:ablation}
\end{table}

The ablation reveals a task-dependent pathway dominance (Table~\ref{tab:ablation}):

\paragraph{Input-delay task (memory retention)}
At $K{=}24$, the xproj-only variant (0.259) achieves comparable performance to the full model (0.667), while bcoup-only (26.47) degrades.
This suggests that the selectivity pathway ($x_{\mathrm{proj}} \to \Delta t, B_t, C_t$) has a greater influence at longer memory horizons.

\paragraph{NARMA-10 (bilinear computation)}
The xproj-only variant ($1.54 {\times} 10^{-3}$) matches the full model exactly, while bcoup-only ($1.96 {\times} 10^{-3}$) captures only part of the benefit.
NARMA-10 requires products ($y_t \cdot \sum y_{t-i}$, $u_{t-9} \cdot u_t$), and the selectivity pathway routes the bilinear product through $\Delta t$, $B_t$, $C_t$ simultaneously, enabling state-dependent gating that implements these multiplicative operations. GM cannot compute such products within the recurrence, and Coupled's dense but linear coupling likewise cannot change the order of computation.

\paragraph{Both pathways are complementary}
On NARMA-10, bcoup-only captures only part of the benefit, confirming that both pathways contribute. On the input-delay task, the selectivity pathway alone suffices at $K{=}24$, but the full model remains competitive across both benchmarks.

\section{Discussion}

\subsection{Computational Cost}

\begin{table*}[t]
\centering
\footnotesize
\setlength{\tabcolsep}{4pt}
\begin{tabular}{@{}lrrrrrrrrr@{}}
\toprule
& & \multicolumn{4}{c}{\textbf{Input-Delay}} & \multicolumn{4}{c}{\textbf{NARMA-10}} \\
\cmidrule(lr){3-6} \cmidrule(lr){7-10}
\textbf{Model} & Params & Train (s) & AR MSE & Inf base (ms) & Inf comp.\ (ms) & Train (s) & AR MSE & Inf base (ms) & Inf comp.\ (ms) \\
\midrule
Standard        & 504 / 312 & 2,601 & 16.99 & 0.33 & 0.31 & 1,568 & $4.22\mathrm{e}{-3}$ & 0.34 & 0.31 \\
Coupled         & 600 / 384 & 2,952 & 27.05 & 1.12 & 0.22 & 2,322 & $3.52\mathrm{e}{-3}$ & 1.09 & 0.21 \\
GM      & 984 / 576 & 2,578 & 0.692 & 1.31 & 0.23 & 2,440 & $3.43\mathrm{e}{-3}$ & 1.27 & 0.22 \\
seq-BIM & 984 / 576 & 3,844 & 0.317 & 9.84 & 0.95 & 3,555 & $1.19\mathrm{e}{-3}$ & 9.74 & 0.93 \\
p-BIM$^{\dagger}$   & 984 / 576 & 2,628 & 0.662 & 2.43 & 0.29 & 2,593 & $1.30\mathrm{e}{-3}$ & 2.42 & 0.29 \\
\bottomrule
\end{tabular}
\caption{Computational cost and inference speed (11 seeds, mean). Models are trained for 200K iterations. Inf base is the per-step autoregressive rollout latency without compilation, and Inf comp.\ is the same metric with \textit{torch.compile} (\textit{reduce-overhead}), measured at sliding window length $50$, median over 200 timed iterations. Input-delay uses $d_{\mathrm{model}}{=}3$, NARMA-10 uses $d_{\mathrm{model}}{=}2$. RTX 3080 Ti, CUDA 12.8, PyTorch 2.10. $\dagger$ p-BIM input-delay AR MSE is averaged over the 10 convergent seeds (Table~\ref{tab:ringbuffer}).}
\label{tab:cost}
\end{table*}

Table~\ref{tab:cost} summarizes computational costs.
We used \textit{torch.compile} (PyTorch 2.10, \textit{reduce-overhead} mode) to optimize model execution by just-in-time (JIT) graph compilation.
GM achieves strong accuracy on the input-delay task with no additional training cost, as its gate modulation is fully parallelizable.
seq-BIM's sequential loop increases training time by approximately $2$--$3\times$, but the accuracy gain per compute cost is favorable on NARMA-10.
At inference, \textit{torch.compile} reduces seq-BIM's latency to 0.93\,ms/step, sufficient for real-time control applications.

p-BIM avoids the sequential loop, bringing training time back to the Std/Coupled/GM range ($\sim\!2{,}600$\,s) while preserving the full matrix-valued bilinear product.
At inference, the per-step autoregressive rollout (Table~\ref{tab:cost}) places p-BIM within the GM and Coupled range ($0.29$\,ms/step compiled), approximately $3\times$ faster than seq-BIM ($0.93$--$0.95$\,ms/step). The matrix-valued $G(x_t)$ introduces only one additional $d_s\!\times\!d_s$ matmul for $N(x_t)$ and one matvec for the state update relative to GM, leaving the per-step cost in the same range.

\subsection{Scaling with $L$ and $d_s$}
\label{app:scaling}

We investigate how the sequence length $L$ and state dimension $d_s$ affect the performance of the bilinear variants (seq-BIM and p-BIM) on NARMA-10, and compare with Standard Mamba, Coupled, and GM under the same conditions.
Results with 11 seeds are reported as medians to reduce outlier sensitivity, while 3-seed results use means.

\paragraph{Architecture separation at $d_s{=}16$}
\label{app:arch_sep}

\begin{table}[t]
\centering
\footnotesize
\begin{tabular}{lrrrr}
\toprule
& & \multicolumn{2}{c}{\textbf{AR MSE}} & \\
\cmidrule(lr){3-4}
Model & Params & $L{=}50$ & $L{=}100$ & Impr. \\
\midrule
Standard & 504 & $3.58\mathrm{e}{-3}$ & $3.17\mathrm{e}{-3}$ & $1.0\times$ \\
Coupled & 664 & $3.39\mathrm{e}{-3}$ & $3.16\mathrm{e}{-3}$ & $0.9\times$ \\
GM & 920 & $3.35\mathrm{e}{-3}$ & $3.27\mathrm{e}{-3}$ & $0.9\times$ \\
seq-BIM & 920 & $5.0\mathrm{e}{-5}$ & $3.3\mathrm{e}{-5}$ & $96\times$ \\
p-BIM   & 920 & $\mathbf{3.9\mathrm{e}{-5}}$ & $\mathbf{2.5\mathrm{e}{-5}}$ & $\mathbf{127\times}$ \\
\bottomrule
\end{tabular}
\caption{Architecture comparison at $d_s{=}16$ on NARMA-10 (median over 11 seeds). Both bilinear variants separate from the non-bilinear baselines by more than two orders of magnitude.}
\label{tab:arch_sep}
\end{table}

Coupling alone provides negligible improvement ($3.39 {\times} 10^{-3}$ vs $3.58 {\times} 10^{-3}$ for Standard), and gate modulation adds no further benefit ($3.35 {\times} 10^{-3}$), as shown in Table~\ref{tab:arch_sep}.
Both bilinear variants achieve nearly two-orders-of-magnitude improvement (seq-BIM $96\times$ at $L{=}100$, p-BIM $127\times$), confirming that the bilinear mechanism is uniquely capable of exploiting an enlarged state space, independent of whether the bilinear product is placed on the input ($x$) or on the state-transition ($A$) side.
At the best configuration ($d_s{=}16$, $L{=}100$), the AR trajectory is visually indistinguishable from the ground truth over 150 autoregressive steps (AR MSE $= 3.3 \times 10^{-5}$ for seq-BIM, $2.5 \times 10^{-5}$ for p-BIM).

\paragraph{Sequence length}
NARMA-10 involves a buffer of the 10 most recent inputs.
We vary $L \in \{25, 50, 75, 100\}$ at both $d_s{=}8$ and $d_s{=}16$. $L{<}25$ tends to cause divergence for all models due to insufficient context for the 10-step NARMA buffer.

\begin{table}[t]
\centering
\footnotesize
\setlength{\tabcolsep}{3pt}
\begin{tabular}{@{}rrrrrr@{}}
\toprule
 & & \textbf{Standard} & \textbf{seq-BIM} & \textbf{p-BIM} & \textbf{Impr.} \\
 & & AR MSE & AR MSE & AR MSE & (p-BIM / Std) \\
\midrule
$L$ & $d_s$ & & & & \\
\midrule
25 & 8   & $7.60\mathrm{e}{-3}$ & $2.44\mathrm{e}{-3}$ & $4.51\mathrm{e}{-3}^{\dagger}$ & $1.7\times$ \\
50 & 8   & $4.13\mathrm{e}{-3}$ & $1.01\mathrm{e}{-3}$ & $1.23\mathrm{e}{-3}$           & $3.4\times$ \\
75 & 8   & $3.52\mathrm{e}{-3}$ & $9.79\mathrm{e}{-4}$ & $9.76\mathrm{e}{-4}$           & $3.6\times$ \\
100 & 8  & $3.29\mathrm{e}{-3}$ & $1.02\mathrm{e}{-3}$ & $9.96\mathrm{e}{-4}$           & $3.3\times$ \\
\midrule
25 & 16  & $6.83\mathrm{e}{-3}$ & $4.05\mathrm{e}{-3}$ & $2.93\mathrm{e}{-3}^{\dagger}$ & $2.3\times$ \\
50 & 16  & $3.58\mathrm{e}{-3}$ & $5.0\mathrm{e}{-5}$  & $3.9\mathrm{e}{-5}$            & $92\times$ \\
75 & 16  & $3.34\mathrm{e}{-3}$ & $6.38\mathrm{e}{-5}$ & $\mathbf{2.4\mathrm{e}{-5}}$    & $\mathbf{139\times}$ \\
100 & 16 & $3.17\mathrm{e}{-3}$ & $\mathbf{3.3\mathrm{e}{-5}}$ & $2.5\mathrm{e}{-5}$ & $127\times$ \\
\bottomrule
\end{tabular}
\caption{Effect of sequence length $L$ on NARMA-10 (median over 11 seeds). $\dagger$ at $L{=}25$ p-BIM training diverges on $3$ of the $11$ seeds, and the reported median is taken over the remaining $8$ convergent seeds.}
\label{tab:l_sweep}
\end{table}

\begin{figure}[t]
\centering
\includegraphics[width=\columnwidth]{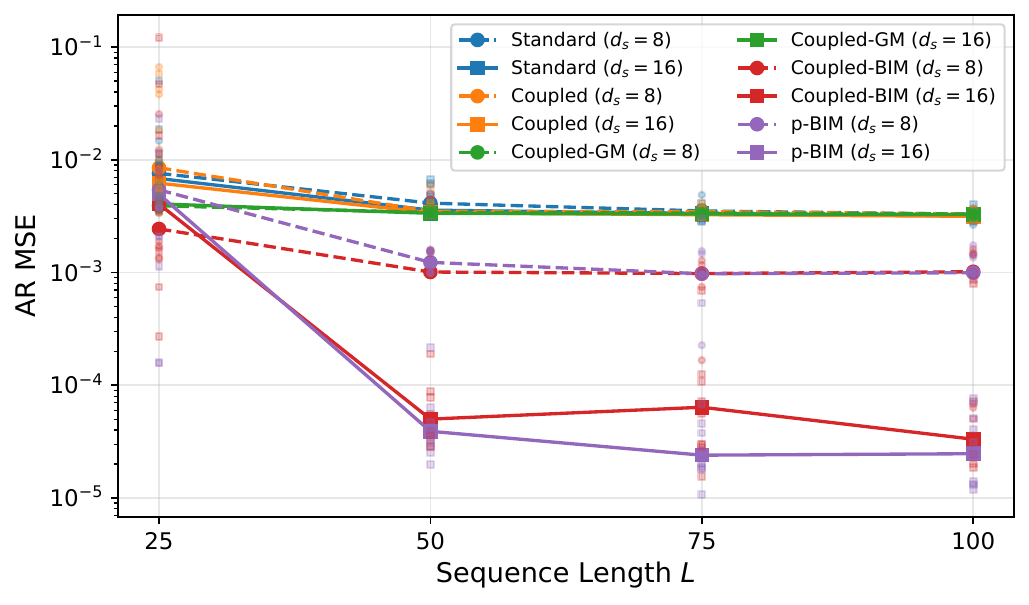}
\caption{NARMA-10: Sequence length $L$ vs AR MSE across architectures at $d_s \in \{8, 16\}$. Dashed: $d_s{=}8$, solid: $d_s{=}16$. Both bilinear variants (seq-BIM and p-BIM) separate clearly from the non-bilinear baselines.}
\label{fig:l_sweep}
\end{figure}

seq-BIM with $d_s{=}8$ improves from $L{=}25$ to $L{=}75$ (AR $9.79 {\times} 10^{-4}$) but saturates at $L{=}100$ ($1.02 {\times} 10^{-3}$), suggesting that $d_i{=}8$ lacks the capacity to exploit longer sequences. With $d_s{=}16$, performance continues to improve up to $L{=}100$ (seq-BIM AR $3.3 {\times} 10^{-5}$, p-BIM AR $2.5 {\times} 10^{-5}$), indicating that the larger state space can absorb more temporal context.
At $L{=}25$ the context is too short for $d_s{=}16$ to help any variant, and 3 of 11 p-BIM seeds diverge (Table~\ref{tab:l_sweep}~$\dagger$).
The divergence is consistent with the lack of an architectural bound on the matrix-valued gate of p-BIM, which allows the coupling matrices to drift during the short warm-up.
The $d_s{=}16$ advantage for the bilinear variants emerges cleanly at $L \geq 50$ and grows with $L$ (Table~\ref{tab:l_sweep}, Fig.~\ref{fig:l_sweep}).

\paragraph{State dimension}
As shown in Table~\ref{tab:l_sweep},
Standard Mamba gains only $1.2\times$ from doubling $d_s$ ($8 \to 16$). Its $d_i {\times} d_s$ independent state variables decay independently, and adding more provides diminishing returns.
seq-BIM gains $20\times$ at $L{=}50$ and $31\times$ at $L{=}100$ from the same change, and p-BIM shows the same qualitative scaling ($32\times$ at $L{=}50$, $40\times$ at $L{=}100$). The bilinear mechanism $(W_x x) \odot \tanh(W_h h)$ (seq-BIM) and its A-side matrix-valued counterpart $N(x_t)$ (p-BIM) both exploit the richer $h$ to compute more complex input-state products.

\begin{figure}[t]
\centering
\includegraphics[width=\columnwidth]{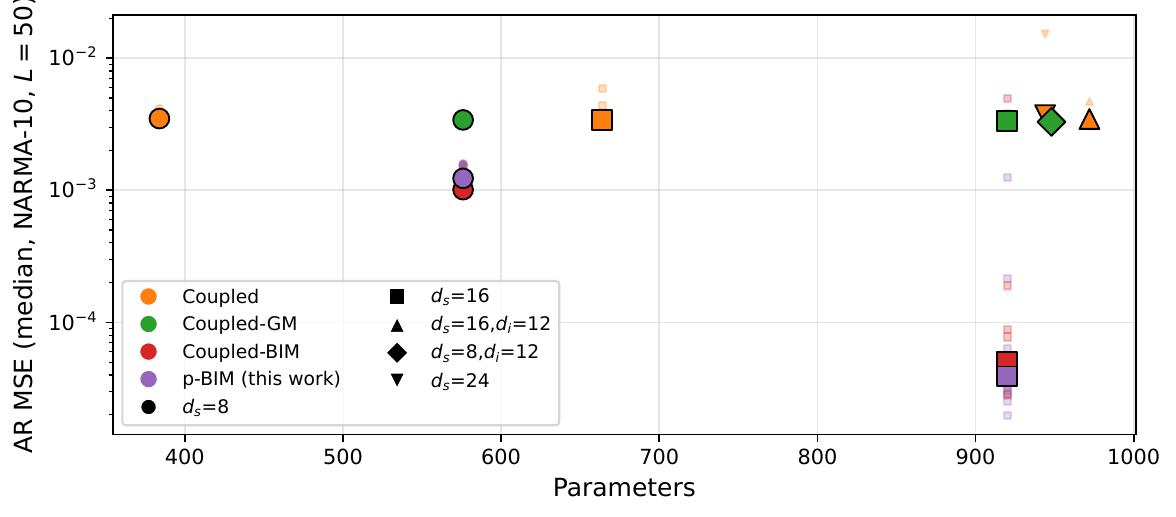}
\caption{Parameter count vs AR MSE (median) on NARMA-10 ($L{=}50$). To ensure a fair comparison at the seq-BIM $d_s{=}16$ parameter count ($\sim$920), we sweep $d_s$ and $d_i$ for Coupled and GM. Standard Mamba is omitted as it did not benefit from increased $d_s$ or $d_i$. All parameter-matched non-bilinear variants cluster near $3{\times}10^{-3}$, while both bilinear variants drop by two orders of magnitude (seq-BIM $5.0{\times}10^{-5}$, p-BIM $3.9{\times}10^{-5}$).}
\label{fig:params_ar_appendix}
\end{figure}

To test whether the improvement is merely a consequence of having more parameters, we match the parameter count of the bilinear variants at $d_s{=}16$ (920 params) for Coupled and GM by varying $d_s$ and $d_i$:
Coupled with $d_s{=}16$, $d_i{=}12$ (972 params) and $d_s{=}24$ (944 params), and GM with $d_s{=}16$ (920 params, exact match) and $d_s{=}8$, $d_i{=}12$ (948 params).
All parameter-matched non-bilinear variants achieve AR MSE in the range $3 {\times} 10^{-3}$ to $8 {\times} 10^{-3}$, indistinguishable from their lower-parameter counterparts.
Both bilinear variants at the same parameter count drop by two orders of magnitude (seq-BIM $5.0 {\times} 10^{-5}$, p-BIM $3.9 {\times} 10^{-5}$, median), confirming that the improvement is due to the bilinear mechanism, not parameter count (Fig.~\ref{fig:params_ar_appendix}).

\subsection{Limitations}
\label{sec:limits}

Our analysis uses small-scale models ($\sim$300--1K parameters) on two benchmarks.
While Section~\ref{app:scaling} demonstrates that the results extend to $d_s{=}16$ with substantial further improvement, scaling to higher-dimensional systems, larger models, and multi-layer architectures remains to be investigated.
The sequential limitation of seq-BIM is partially addressed by p-BIM, which recovers parallel scanning at matched training cost and matches the GM/Coupled per-step rollout latency under \textit{torch.compile} (Table~\ref{tab:cost}). Reducing the across-seed variance of p-BIM on the input-delay task remains open.

\section{Conclusion}

We proposed Mamba-X, a bilinear input modulation for selective SSMs inspired by the Koopman bilinear form, and showed that it cleanly separates into two mechanisms, namely gate modulation for memory retention (GM) and bilinear computation via state-input products (seq-BIM).
The scaling analysis reveals that only the bilinear mechanism can exploit larger state spaces, with seq-BIM achieving $96\times$ improvement at $d_s{=}16$ while all other variants remain flat. This reveals a qualitative, not merely quantitative, distinction.
A parallel-scannable realisation of the same bilinear mechanism (p-BIM, on the $A$-side) matches seq-BIM's scaling at lower training cost, partially answering the open question of whether the bilinear capacity can be recovered without sequential computation.

\bibliographystyle{IEEEtran}
\bibliography{references}

\end{document}